\documentclass[usegraphicx]{mn2e}
\usepackage{amsmath}
\usepackage{amssymb}
\usepackage{times}
\usepackage{subfigure}
\usepackage{epsfig}

\renewcommand{\descriptionlabel}[1]%
  {\hspace{\labelsep}\textbf{#1}}

\title[Blazhko variables in NGC 5024]{The unusually large population of Blazhko variables in the globular cluster 
NGC 5024 (M53)\thanks{Based on observations collected at the Indian
   Astrophysical Observatory, Hanle, India.}}

\author[A. Arellano Ferro et al.]
{A. Arellano Ferro$^{1}$\thanks{E-mail:armando@astro.unam.mx},  
D.M. Bramich$^{2}$\thanks{E-mail:dan.bramich@hotmail.co.uk},
R. Figuera Jaimes$^{1}$\thanks{E-mail:rfiguera@astro.unam.mx},  
Sunetra Giridhar$^{3}$\thanks{E-mail:giridhar@iia.res.in},
\and
K. Kuppuswamy$^{3}$\thanks{E-mail:kuppuswamy@iia.res.in}\\
  \medskip
\\$^{1}$Instituto de Astronom\1a, Universidad Nacional Aut\'onoma de M\'exico\\
\\$^{2}$European Southern Observatory, Karl-Schwarzschild-Stra$\beta$e 2, 85748
Garching bei M\"{u}nchen, Germany\\
\\$^{3}$Indian Institute of Astrophysics, Koramangala 560034, Bangalore, India
}

\begin{document}
 
\date{Accepted . Received ; in original form }

\pagerange{\pageref{firstpage}--\pageref{lastpage}} \pubyear{2011}

\maketitle 

\label{firstpage}

\begin{abstract}
We report the discovery of amplitude and phase modulations typical of the Blazhko effect in 
22 RRc and 9 RRab type RR Lyrae stars in NGC 5024 (M53). 
This brings the confirmed Blazhko variables in this cluster to 23 RRc and 11 RRab, that represent 66\% and 37\% 
of the total population of RRc and RRab stars in the cluster respectively,  making NGC 5024 the globular cluster with the largest presently known population of Blazhko RRc stars. We place a lower limit on the overall
incidence rate of the Blazhko effect among the RR Lyrae population in this cluster of 52\%. New data have allowed us to refine the pulsation periods. The limitations imposed by the time span and sampling of our data prevents reliable
estimations of the modulation periods. The amplitudes of the modulations range between 0.02 and 
0.39 mag. The RRab and RRc are neatly separated in the CMD, and the RRc Blazhko variables are on averge redder than their stable couterparts; these two facts
may support the hypothesis that the HB evolution in this cluster is towards
the red and that the
Blazhko modulations in the RRc stars are connected with the pulsation mode switch.
\end{abstract}
      

\section{Introduction}

In 1907, S. Bla$\check{\rm z}$ko described 
the cyclic variation of the pulsation period of RW Dra, without comments on the amplitude behaviour, 
(Bla$\check{\rm z}$ko 1907), and in 1916, H. Shapley reported cyclic variations in the shape of the light curve, period and amplitude in the star RR Lyrae (Shapley 1916). Period oscillation and/or amplitude modulation in RR Lyrae type stars 
is generally referred to as the  Bla$\check{\rm z}$ko or Blazhko effect.
Over the last hundred years numerous RR Lyrae stars have been found to exhibit cyclic modulations of both period and amplitude although cases of amplitude modulations with no period variations are common. The cause of these modulations has remained unsatisfactorily explained on theoretical grounds.
This is not surprising since, to detect and characterize the variations, even in 
the cases with large amplitude modulations, large sets of accurate photometric observations obtained over a long time span are required. 
Most recently, as a result of careful inspection of large data bases (e.g. Moskalik \& Poretti et al. 
2003 (OGLE); Kolenberg et al. 2010 ($Kepler$)), the detection of the Blazhko effect in RR Lyrae stars has 
improved. Long term surveys have proven useful
in the detection of very small amplitude modulations, increasing the fraction of known Blazhko variables among 
the fundamental pulsators RRab (RR0) to about 50\% (e.g. the Konkoly Blazhko Survey; Jurcsik et al. 2009) 
and have also allowed detailed studies of individual objects (e.g. Chadid et al. 2010a (V1127 Aql); Chadid et al. 2010b (S Arae);
Poretti et al. 2010 (CoRoT 101128793); Kolenberg et al. 2011 (RR Lyr)).

\begin{table*}
\caption{Time-series $V$ and $I$ photometry for all the RR Lyrae stars in our field of view.
         The standard $M_{\mbox{\scriptsize std}}$ and instrumental $m_{\mbox{\scriptsize ins}}$ magnitudes are listed in columns 4 and 5, respectively, corresponding to the variable
         star, filter, and epoch of mid-exposure listed in columns 1-3, respectively. The uncertainty on $m_{\mbox{\scriptsize ins}}$ is listed
         in column 6, which also corresponds to the uncertainty on $M_{\mbox{\scriptsize std}}$. For completeness, we also list the quantities $f_{\mbox{\scriptsize ref}}$, $f_{\mbox{\scriptsize diff}}$
         and $p$ (see Eq. 1 in Paper I) in columns 7, 9 and 11, along with the uncertainties $\sigma_{\mbox{\scriptsize ref}}$ and $\sigma_{\mbox{\scriptsize diff}}$
         in columns 8 and 10.
         This is an extract from the full table, which is available with the electronic version of the article (see Supporting Information).
         }
\centering
\begin{tabular}{ccccccccccc}
\hline
Variable & Filter & HJD & $M_{\mbox{\scriptsize std}}$ & $m_{\mbox{\scriptsize ins}}$ & $\sigma_{m}$ & $f_{\mbox{\scriptsize ref}}$ & $\sigma_{\mbox{\scriptsize ref}}$ & $f_{\mbox{\scriptsize diff}}$ &
$\sigma_{\mbox{\scriptsize diff}}$ & $p$ \\
Star ID  &        & (d) & (mag)                        & (mag)                        & (mag)        & (ADU s$^{-1}$)               & (ADU s$^{-1}$)                    & (ADU s$^{-1}$)                &
(ADU s$^{-1}$)                     &     \\
\hline
V1 & $V$ & 2454939.20002    & 17.167 & 18.487 & 0.003  & 661.567 & 1.722  & -299.031 & 1.098  & 1.1564 \\
V1 & $V$ & 2454939.23093    & 17.187 & 18.507 & 0.003  & 661.567 & 1.722  & -309.195 & 1.076  & 1.1625 \\
\vdots   & \vdots & \vdots  & \vdots & \vdots & \vdots & \vdots   & \vdots & \vdots   & \vdots & \vdots \\
V1 & $I$ & 2454939.19220    &16.541  & 17.935 & 0.003  & 782.781  & 3.855  &-134.485  & 2.042 & 1.1907 \\
V1 & $I$ & 2454939.20744    &16.528  & 17.922 & 0.003  & 782.781  & 3.855  &-122.903  & 1.961 & 1.1757 \\
\vdots   & \vdots & \vdots  & \vdots & \vdots & \vdots & \vdots   & \vdots & \vdots   & \vdots & \vdots \\
\hline
\end{tabular}
\label{tab:vri_phot}
\end{table*}
 
While knowledge of the Blazhko population in globular clusters would yield important insight to the 
metallicity influence on
the incidence of the Blazhko phenomenon, the study of Blazhko variables in globular clusters is still 
very much unexploited.
Several difficulties including that stars are faint and often in very crowded fields mean that long
time series of high quality CCD images and photometry of exceptional high quality are required.

During the course of an investigation into the variable stars in the globular cluster NGC 5024 (M53) based on $V$ and $I$ CCD photometry (Arellano Ferro et al. 2011; Paper I) , clear amplitude and light curve shape variations were detected in a few RRab stars and in the majority of the RRc stars. 
These Blazhko-like variations are surprising since the Blazhko effect has been mostly associated with 
RRab stars and not so much with RRc stars. Amplitude and phase 
modulations were found in RRc stars only recently in three stars in M55 (namely V9, V10 and V12) 
(Olech et al. 1999), in the LMC (2-4\% of the RRc stars) (Alcock et al. 2000), in NGC 6362 (namely V6, V10 and V37) (Olech at al. 2001) and in three RRc stars in the OGLE database (BWC V47, MM5A V20 and BW8 V34) 
(Moskalik \& Poretti 2003). In these stars, non-radial modes with very close frequencies to the primary 
frequency have been identified and the beating of these pulsational modes results in long-term amplitude and phase modulations. A different scenario is
that the modulation produces asymetric sidepeaks to the primary but that due
to limited accuracy of ground-based data one only sees the most prominent peaks
(Benk\H o et al. 2010).
However, in higher accuracy space-based data, such as from the CoRoT and Kepler missions, the full spectrum of side peaks is always visible irrespective of what the cause of the modulations might be.

In this paper we present the detection of Blazhko-type amplitude and phase modulation in the majority of RRab and RRc variables in the globular cluster NGC~5024.
Despite the limited time span of our data and
of the sparse coverage of the modulation cycle, we have made an attempt
to estimate the modulation period, which we find to be reliable only in a few 
cases.

The globular cluster NGC 5024 (M53) (R.A.(2000)$13^h 12^m \,55^s. 3$, DEC(2000)$=+18^{\circ} 10'09''$) is located in the intermediate galactic halo  (l=332.97, b=+79.77, $Z=$ 17.5 kpc, R$_G$=18.3 kpc and as a consequence it is subject to very low reddening ($E(B-V) = 0.02$). Its metallicity [Fe/H]~$\sim -1.92 \pm 0.06$ and an average distance d$\sim 18.3 \pm 0.4$ kpc have been calculated from the RR Lyrae stars in Paper I.

\section{Observations and Reductions}
\label{sec:Observations}

We have complemented the observations reported in Paper I with new Johnson $V$ and $I$ 
observations obtained on March 12 and 13, and April 11, 12, 13 and 14, 2011. 
The observations were performed with
the 2.0m telescope of the Indian Astronomical Observatory (IAO), Hanle, India, located at 4500m above sea level. The estimated seeing was $\sim$1 arcsec.
The detector was a 
Thompson CCD of 2048 $\times$ 2048 pixels with a pixel
scale of 0.296 arcsec/pix and a field of view 
of approximately $10.1 \times 10.1$ arcmin$^2$.
The new observations have brought the number of available epochs from 
177 in the $V$-band and 184 in the $I$-band in Paper I, to 297 and 307 in $V$ and $I$
respectively.
As in Paper I, we employed the technique of difference image analysis (DIA) to extract high precision photometry
for all point sources in the images of NGC 5024 (Alard \& Lupton 1998; Alard 2000;
Bramich et al. 2005).
We used a pre-release version of the {\tt DanDIA}\footnote{
{\tt DanDIA} is built from the DanIDL library of IDL routines available at http://www.danidl.co.uk}
pipeline for the data reduction process (Bramich et al., in preparation) which includes a new algorithm
that models the convolution kernel matching the point-spread function (PSF) of a pair of images of the same field as a discrete pixel array
(Bramich 2008). The transformations to the standard system were as
discussed in Paper I. For the sake of homogeneity we choose to use the same transformation equations calculated in Paper I to transform the new set of data
to the standard Johnson-Kron-Cousins photometric system.
All our new $V,I$ photometry for the variables in the field of our images for NGC 5024  is reported in Table \ref{tab:vri_phot}. For the sake of
commodity to the interested reader we have also included in the table the
$V,I$ data from Paper I supplemented by the new observations. The parameters
$f_{\mbox{\scriptsize ref}}$, $f_{\mbox{\scriptsize diff}}$ and $p$ are 
fundamental to the difference imaging method and allow to reconstruct the original fluxes on the images, hence they are included in Table \ref{tab:vri_phot}. Their exact definitions are given in Paper I (section 2) and the 
interested reader is refered to that paper.
Only a small portion of Table \ref{tab:vri_phot} is given in the printed version of this paper but the full table is available in electronic form.

\begin{table*}
\caption{Periods and epochs of the RR Lyrae stars in NGC~5024 in our field of view.}
 \begin{tabular}{llcc||llcc}
\hline
 Variable & Bailey's & $P$ & HJD$_{max}$ &Variable & Bailey's & $P$ & HJD$_{max}$ \\
          & type     & (days) &(+240~0000.)&  & type     & (days) &(+240~0000.) \\
\hline
V1 & RRab&0.609823&55633.427&V38& RRab-Bl&0.705793&55220.482 \\
V2 & RRc-Bl&0.386146&54940.342&V40& RRc&0.314820&55264.433 \\
V3 & RRab&0.630596&55633.450&V41& RRab-Bl&0.614440&55323.107 \\
V4 & RRc-Bl&0.385609&55666.170&V42& RRab&0.713716&55665.270\\
V5 & RRab&0.639425&55249.474&V43& RRab-Bl&0.712014&54941.204 \\
V6 & RRab&0.664019&55294.248&V44& RRc-Bl&0.374908&55666.237\\
V7 & RRab&0.544860&54940.391&V45& RRab&0.654944&55633.439\\
V8 & RRab&0.615524&55666.180&V46& RRab-Bl&0.703649&55665.230\\
V9 & RRab&0.600371&55633.411&V47& RRc-Bl&0.335366&55249.380 \\
V10& RRab&0.608267&55665.264&V51& RRc-Bl&0.355215&55663.253\\
V11& RRab-Bl&0.629946&55665.270&V52& RRc-Bl$^a$&0.364621&55666.242\\
V15& RRc-Bl&0.308692&55666.242&V53& RRc-Bl$^a$&0.389080&55633.411\\
V16& RRc-Bl&0.303160&55294.257 &V54& RRc-Bl&0.315107&55664.379\\
V17& RRc-Bl&0.381098&55323.183&V55& RRc-Bl&0.443227&55666.252\\
V18& RRc-Bl&0.336060&55666.242&V56& RRc&0.328894&55220.457 \\
V19& RRc&0.391159&55633.387&V57& RRab-Bl&0.568244&55633.379\\
V23& RRc-Bl&0.366141&54940.391 &V58& RRc-Bl&0.354971&55663.244\\
V24& RRab&0.763201&55666.229&V59& RRc-Bl&0.303942&55664.334\\
V25& RRab-Bl&0.705153&55264.341 &V60& RRab-Bl&0.644756&55633.368\\
V27& RRab&0.671067&54940.342 &V61& RRc-Bl&0.369930&54940.176\\
V29& RRab-Bl&0.823249&55634.507&V62& RRc-Bl&0.359887&55663.261 \\
V31& RRab-Bl&0.705670&55634.386&V63& RRc-Bl&0.310466&55634.403\\
V32& RRc-Bl&0.390076&55634.393&V64& RRc-Bl&0.319724&55634.378\\
V33& RRab-Bl&0.624589&54941.416 &V71& RRc&0.304509&54939.249 \\
V34& RRc&0.289625&54941.266&V72& RRc&0.340749&55664.370\\
V35& RRc-Bl&0.372668&55264.280 &V91& RRc&0.302426&55634.474\\
V36& RRc-Bl&0.373301&54941.430&V92& RRc&0.277222&55633.509 \\
V37& RRab&0.717617&55663.249& &  & &  \\
\hline
\end{tabular}

Bl: Blazhko variables with the $\cal S_B$-{\it index} larger than 145. $a$: V52 and
V53 lie in close spatial proximity and it is possible that flux contamination between the variables is, at least a partial cause of the observed Blazhko effect.
\label{variables}
\end{table*}
 
\section{Pulsation Periods}
\label{sec:pul_periods}

We have applied the string-length method (Burke et al. 1970; Dworetsky 1983) to the light curves which determines the best period and a corresponding normalized 
string-length statistic $S_{\mbox{\scriptsize Q}}$ to estimate
the pulsation periods of all the RR Lyrae stars in our sample. The so determined periods and the epochs employed to phase the light curves are listed in Table \ref{variables}.

\section{Light curves of the RR Lyrae stars}
\label{sec:LIGHTCURVES}

\begin{figure*}
\begin{center}
\includegraphics[width=18.cm,height=20.cm]{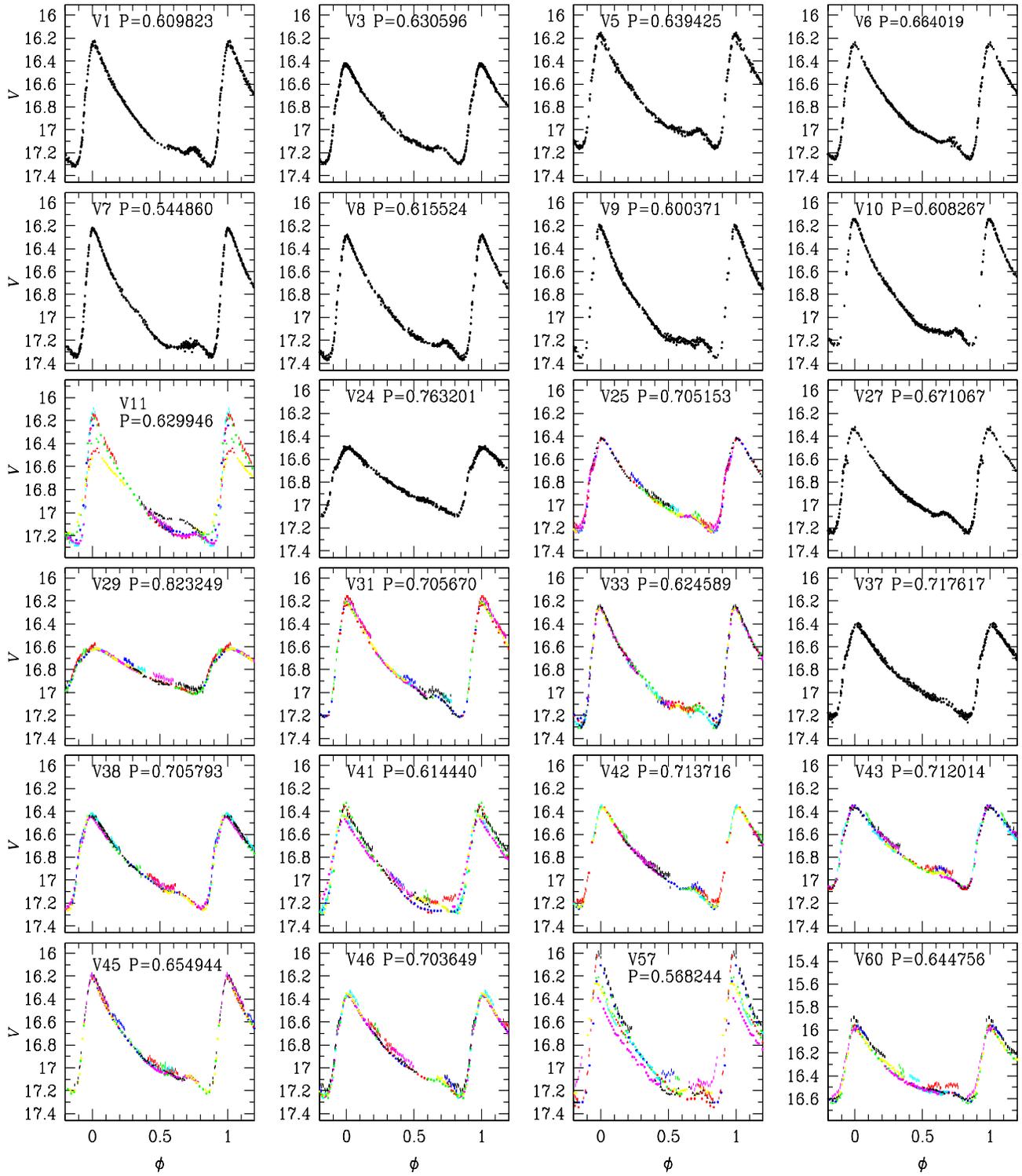}
\caption{Phased $V$ light curves of the RRab stars. For stars showing signs of amplitude and phase 
modulations, different colours have been used for individual 
nights following the code given in Table \ref{tab:colourcode}. Stars for which the light curves are seemingly
non-modulated are displayed with black dots and give a sense of the reliability of our photometry.}
    \label{BLAZHKO_RR0_V}
\end{center}
\end{figure*}

\begin{figure*}
\begin{center}
\includegraphics[width=18.cm,height=20.cm]{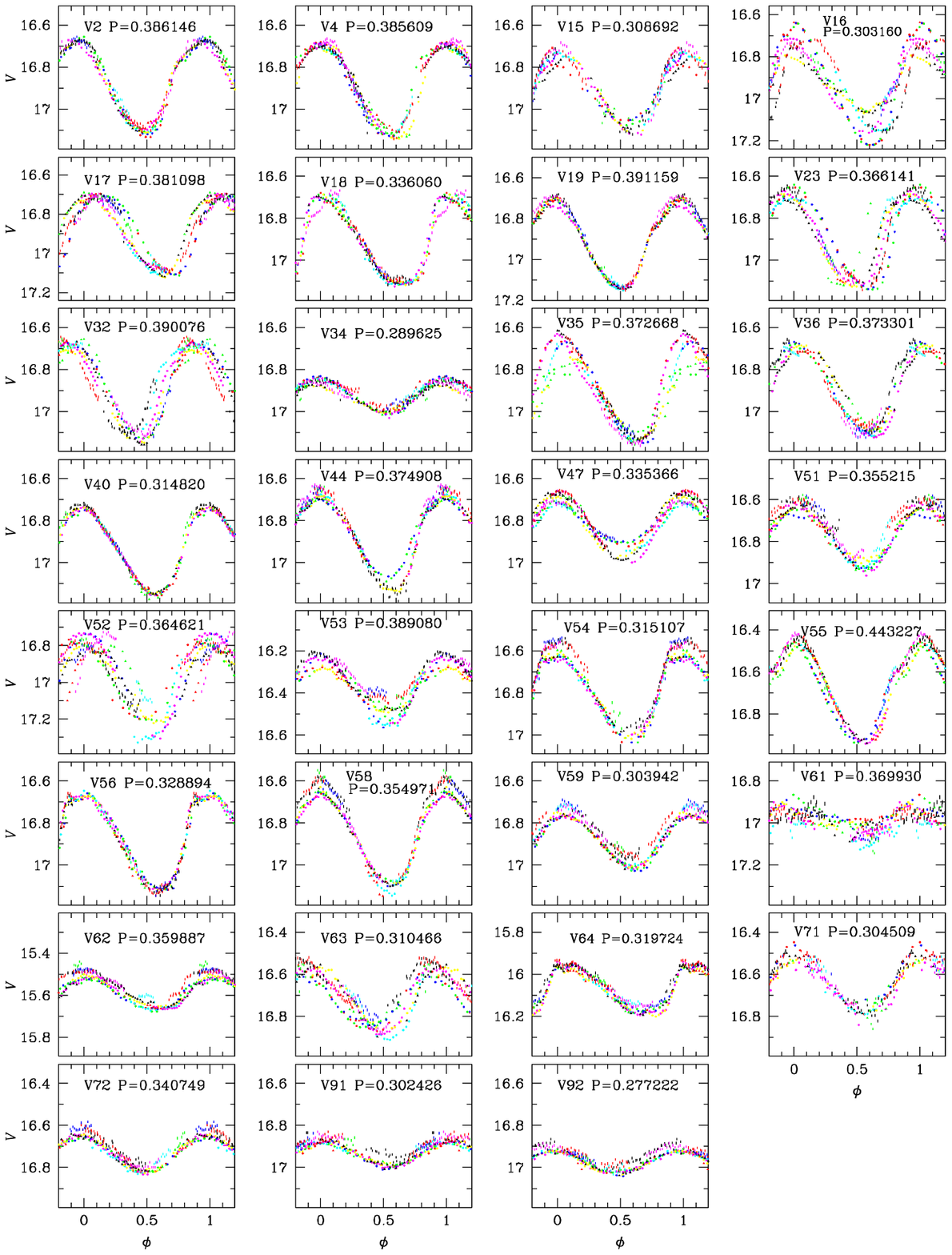}
\caption{Phased $V$ light curves of the RRc stars. The colour-code for the symbols is the same as in 
Fig. \ref{BLAZHKO_RR0_V} (as defined in Table \ref{tab:colourcode}).}
    \label{BLAZHKO_RR1_V}
\end{center}
\end{figure*}

The $V$ light curves of the RRab and RRc stars, phased with the ephemerides in Table \ref{variables}, are displayed in Figs. \ref{BLAZHKO_RR0_V} and 
\ref{BLAZHKO_RR1_V}. The phase coverage is very good in all cases.
Different colours and symbols have been used to
highlight the nightly variations of the light curve shape. The date-colour code is listed 
in Table \ref{tab:colourcode}. The night of January 22 (JD 2455219.) only contains two data points, hence it was
neither coded nor plotted.

Amplitude and phase modulations are clearly visible in the light curves for most stars. Light curves without visible evidence of modulations have been plotted with black
symbols and it can be argued that the stability of these stars helps in proving that
the modulations detected in the rest of the stars are real and not an artifact of our reductions. Further insight on the stability of the reduction process will be addressed in section \ref{sec:syserr}.

The light curves in the $I$-band have been carefully explored and 
despite having smaller amplitude variations than in the $V$-band,
and hence smaller Blazhko modulations, the effect is very obvious and
consistent with the $V$ band counterparts. For the sake of brevity, we do not plot the $I$ light curves but the data are included in 
Table \ref{tab:vri_phot}.



\begin{figure*} 
\includegraphics[width=18.cm,height=20.cm]{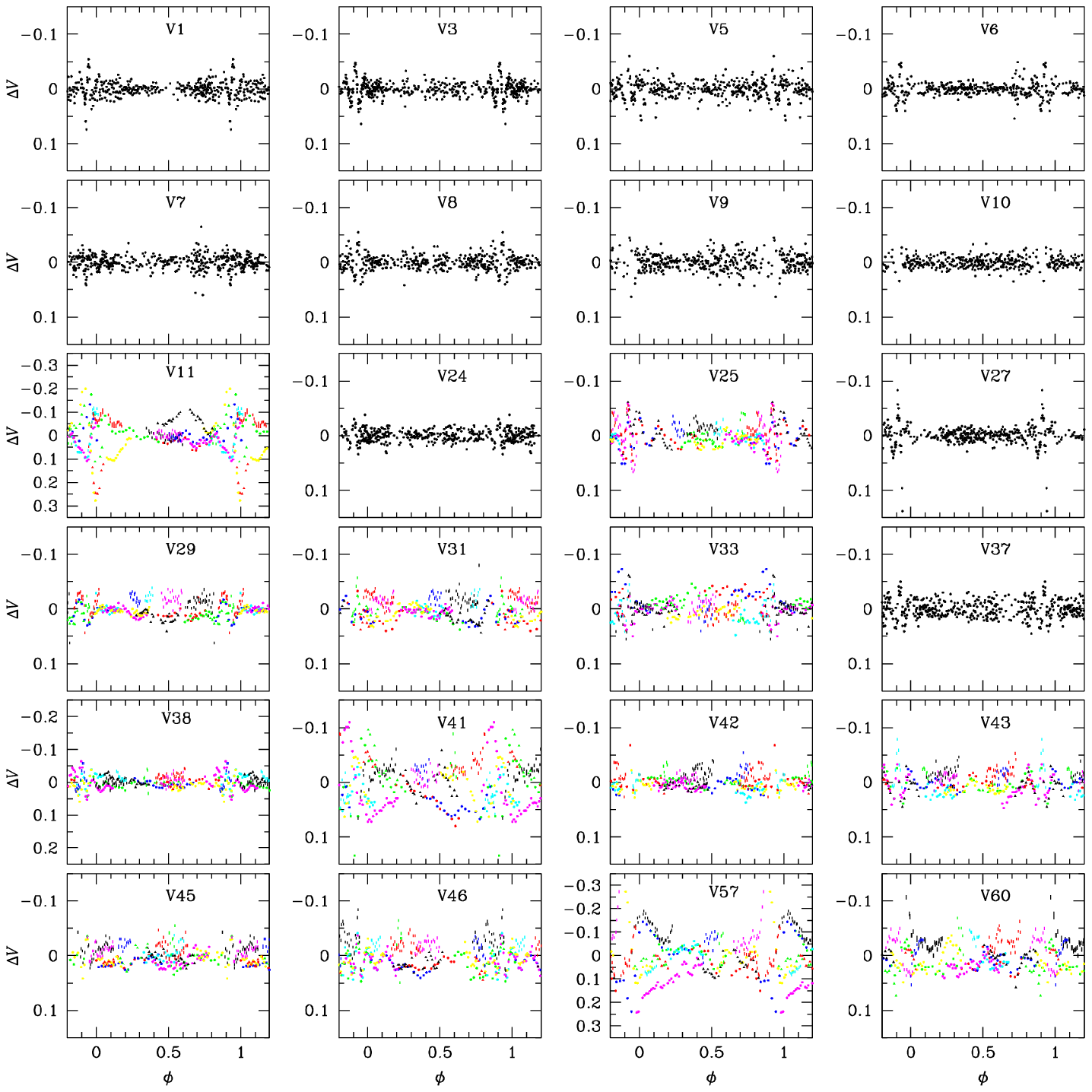}
\caption{Residuals of the $V$ phased light curves relative to a fitted Fourier model for the RRab stars. Systematic nightly differences relative to 
the model phased light variation can be clearly observed 
in most variables which we interpret as due to the Blazhko amplitude and/or phase modulation. See the text for a discussion on individual cases.
The colour code is as in Fig. \ref{BLAZHKO_RR0_V}. Note that the vertical scale is  different for a few stars with the largest amplitude modulations (V11, V38 and V57). 
Typical photometric uncertainty is 0.008 mag.}
    \label{DIFS_RR0}
\end{figure*}

\begin{figure*} 
\includegraphics[width=18.cm,height=20.cm]{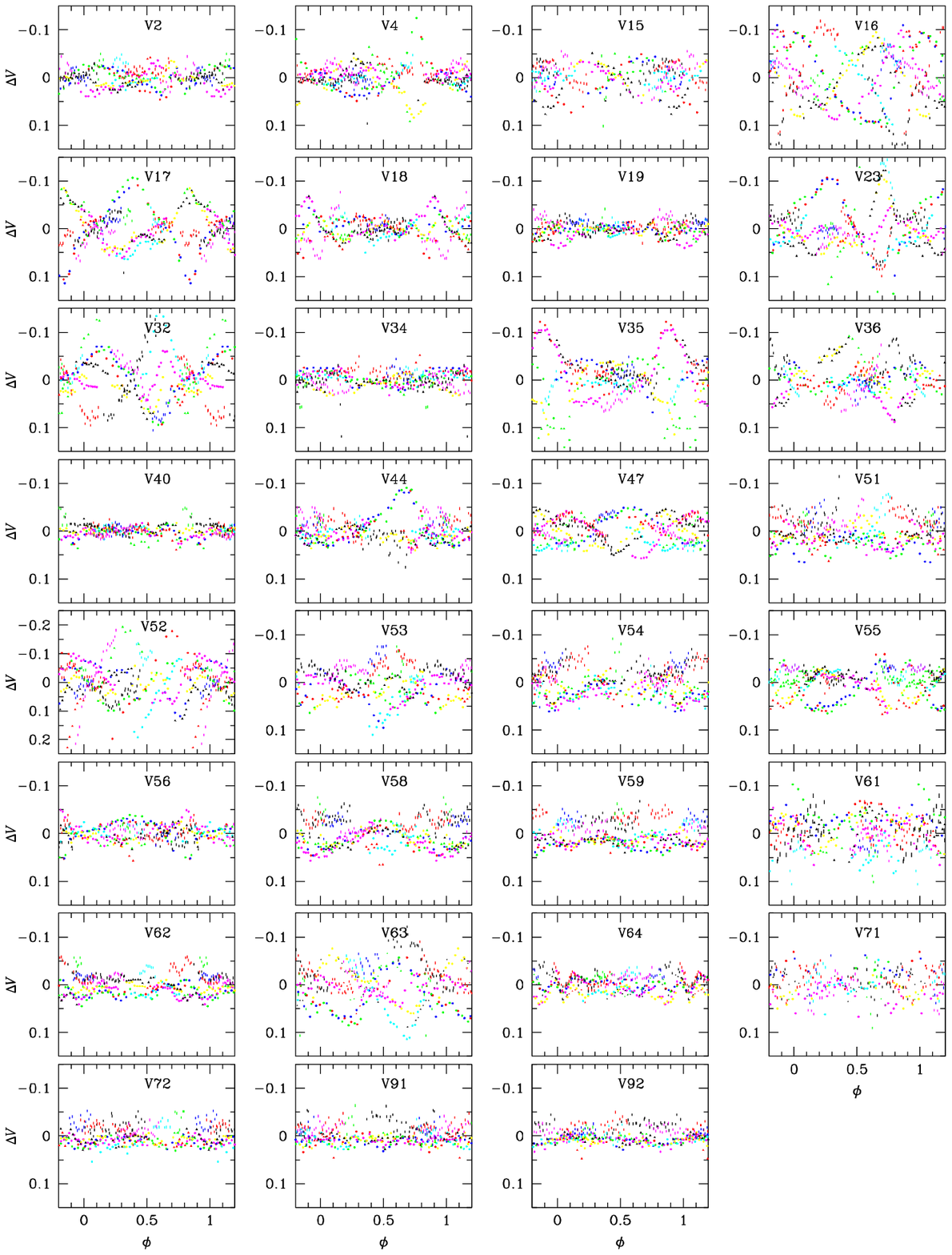}
\caption{Residuals of the $V$ phased light curves relative to a fitted Fourier model for the RRc stars.
See also the caption to Fig.  \ref{DIFS_RR0}.}
    \label{DIFS_RR1}
\end{figure*}

\section{RR Lyrae stars with the Blazhko effect}
\label{sec:Blazhko}

RR Lyrae stars exhibiting the Blazhko effect in NGC~5024 have been reported by D\'ek\'any \& Kov\'acs (2009); 
two RRab stars, V11 and V57, and one RRc star, V16. The corresponding light curves in 
Fig. \ref{BLAZHKO_RR0_V} and \ref{BLAZHKO_RR1_V} indeed exhibit clear signs of the amplitude modulation. We have noticed 
amplitude and light curve shape variations in the majority of the RRc and in a number
of other RRab stars. 
The appearance of the light curves in most RRc stars resembles the light curves of the amplitude 
modulated RRc stars found by Olech et al. (1999) in M55. 
Clear amplitude changes are observed over time scales of days. In order to make
the time variations of the light curves more evident, in Figs. \ref{DIFS_RR0} 
and \ref{DIFS_RR1} we plot the residuals of the phased light curves relative to a fitted Fourier
model (see Paper I; fits are redone including the new photometric data) following
the same colour code in Table \ref{tab:colourcode}. These plots 
highlight the amplitude variation scale and 
help to distinguish small, rather marginal, amplitude modulations. 

We inspected each star in the reference image
looking for blends or other possible defects that may cause systematic errors in the overall light curve amplitudes. 
A blended reference flux will not change the shape of the light curve but it will affect the amplitude (see Bramich et al. 2011
for a more detailed discussion). 
Hence, any Blazhko variations will exhibit exactly the same form regardless of the reference flux, 
but the measured amplitudes of these excursions from the mean curve may be affected.
Then, it is convenient to note the 
location of each variable in the general field of the cluster. We offer a guide of the 
level of blending by defining the ``inner'' and the ``outer'' regions of the cluster. 
Any star in the central region of the cluster is most likely blended with fainter stars 
whereas in the outer region this problem is negligible.
Despite the good PSF subtraction of most stars in the cluster when measuring the reference fluxes on the
reference image, in the central region the 
systematic residuals are always more prominent than in the outer region. An examination of the 
residuals of the reference image suggests as a reasonable border-line a circle of radius 60~arcsec 
centred on the cluster. In the comments column of Tables \ref{tab:blazhko} and \ref{tab:nonBl}, we indicate to 
which region each star belongs and the extent of any blending. 
While the brightness of some stars is indeed very likely 
contaminated by nearby stars, some others are comfortably isolated. 
We decided to retain the complete sample for further evaluation.

At this stage, we deem it necessary to demonstrate that the observed
amplitude and phase modulations of the RR Lyrae stars in NGC~5024 are not caused by inaccuracies in
our observations and/or systematic errors introduced by our reduction software. The following subsections deal with
this aspect.

\begin{table}
\caption{Julian date colour-code used in Figs. \ref{BLAZHKO_RR0_V}, \ref{BLAZHKO_RR1_V},
\ref{DIFS_RR0}, \ref{DIFS_RR1} and \ref{fakelc}.}
         \begin{center}
\begin{tabular}{@{}cc@{}}
\hline
HJD (d) & Colour   \\

\hline
245 4939.& red circles\\
245 4940.& green circles\\
245 4941.& blue circles\\
245 5220.& turquoise circles\\
245 5249.& purple circles\\
245 5263.& yellow circles\\
245 5264.& black triangles\\
245 5294.& red triangles\\
245 5323.& green triangles \\
245 5633.& black tick\\
245 5634.& red tick\\
245 5663.& green tick\\
245 5664.& blue tick\\
245 5665.& turquoise tick\\
245 5666.& purple tick\\
\hline
\end{tabular}
\end{center}
\label{tab:colourcode}
\end{table}

\subsection{Analysis of systematic errors in the photometry}
\label{sec:syserr}

In this section we attempt to disentangle the systematic errors
caused during the {\tt DanDIA} reduction process from the genuine "Blazhko" variations.
Our approach consists of generate simulated time-series calibrated imaging data that contains
simulated RR Lyrae variables with perfectly stable model light curves (i.e. no Blazhko effect),
and then reducing the simulated imaging data with the {\tt DanDIA} pipeline. The extracted
light curves of the simulated RR Lyrae stars will then allow an assessment of the level to 
which systematic errors in the reductions can imitate the Blazhko effect.
Note that we only consider the $V$ filter in this analysis because this is the waveband
in which the Blazhko effect has the greatest photometric amplitude.

\subsubsection{Simulating images of NGC 5024}
\label{sec:simim}

The reductions of the real imaging data of NGC 5024 provide us with all the information
we require to generate a set of time-series model calibrated images for the cluster.
By taking the extra step of adding noise to the model image pixel values using the
adopted noise model for the CCD, we may generate a set of realistic simulated observations
of the cluster closely matching the properties of our real calibrated imaging data.

We start with the list of 9075 stars detected on the reference image for the $V$ filter,
adopting their detected positions and measured reference fluxes as model values. Included in
this list are the 24 RR0 and 31 RR1 variable stars in NGC 5024 for which we have real
light curve data. For each of these variable stars,
we adopt the fitted Fourier decomposition of the phased light curve
as the model light curve, which we convert to flux
units on the reference image using the model reference flux. For all the remaining stars,
we assume that they are non-variables, and we adopt a model light curve consisting of a
constant flux on the reference image equal to the model reference flux.

For each real image in our time-series imaging data,
we carry out the following procedure to generate a corresponding simulated calibrated image of the cluster:
\begin{enumerate}
\item We create a model image array with the same dimensions as a real calibrated image such
      that each pixel value is equal to the measured sky level (ADU) in the current real
      image.
\item We calculate the model flux of each star at the epoch corresponding to the
      current real image using its model light curve,
      and we then adjust this flux by multiplying by the photometric scale factor derived
      for the real image during the image subtraction stage of {\tt DanDIA}.
      This yields a model flux (ADU) for each star on the flux scale of the current real image.
\item We calculate the pixel coordinates of each star in the model image by transforming
      the reference-image pixel-coordinates using the coordinate transformation
      derived between the reference image and the current real image
      during the image registration stage of {\tt DanDIA}.
\item For each star, we generate a model PSF on the model image using the following method.
      We calculate the star PSF on the reference image by evaluating the
      reference-image model-PSF from the {\tt DanDIA} reductions at the reference-image
      pixel-coordinates of the star. The reference-image model-PSF corresponds
      to the pixel centres, and it is therefore necessary to shift the star PSF with
      a subpixel shift (via cubic O-MOMS resampling, see Blu et al. 2001) so as to match the
      subpixel coordinates of the star in the model image. Since any rotation, shear or scale
      change between the real images in our time-series is negligible, no further coordinate
      transformation of the star PSF is necessary. We then convolve the star PSF with the
      kernel solution for the current real image from the {\tt DanDIA} reductions evaluated
      at the reference-image pixel-coordinates of the star. Finally, we enforce the
      star PSF normalisation to a sum of unity.
\item For each star, we add the model PSF generated in step~(iv) scaled by the corresponding
      model flux calculated in step~(ii) to the model image created in step~(i) at the star's
      pixel coordinates calculated in step~(iii). This results in the production of a
      model calibrated image corresponding to the current real image.
\item For each pixel in the model image, indexed by column $i$ and row $j$, we calculate the
      following noise map $\sigma_{ij}$ using the standard CCD noise model for calibrated images
      adopted by the {\tt DanDIA} software:
      \begin{equation}
      \sigma_{ij} = \sqrt{ \frac{\sigma^{2}_{0}}{F^{2}_{ij}} + \frac{M_{ij}}{G F_{ij}} }
      \label{eqn:noise_model}
      \end{equation}
      where $\sigma_{0} =$6.5~ADU is the CCD readout noise, $G =$1.3~e$^{-}$/ADU is
      the CCD gain, $F_{ij}$ is the master flat-field image used to calibrate the current real
      image, and $M_{ij}$ is the model image we have generated in step~(v) (see Bramich 2008).
\item Again, for each pixel in the model image, we generate a random number drawn from a
      normal distribution with zero mean and unit sigma, multiply these numbers by
      the noise map $\sigma_{ij}$, and add them to the model image from step~(v).
      This results in the production of a simulated noisy calibrated image corresponding to
      the current real image.
\item We use a copy of the bad pixel mask for the current real calibrated image as the bad pixel mask
      for the simulated calibrated image, which includes bad pixels corresponding to saturated pixels.
\end{enumerate}
We note that saturated stars are not included in the reference-image star-list that
is provided by the {\tt DanDIA} software, and therefore the simulated images generated by the above
procedure do not include any flux from saturated stars in the reference image. However,
the adoption of the real bad pixel masks for the simulated images ensures that any pixels that could contain
flux from a saturated star are properly masked during the reduction process.

\subsubsection{Reduction of the simulated images of NGC 5024}
\label{sec:red_fake}

We reduced the simulated images generated in Section~\ref{sec:simim} using the 
{\tt DanDIA} pipeline with exactly the same parameters as those used for reducing our real
time-series imaging data. We constructed the simulated reference image from the simulated images corresponding to
the real images used to construct the real reference image, and we detected 8982 stars on the simulated reference image
from the 9075 real star records that were used to generate the simulated images. The pipeline
successfully processed all 297 simulated images, and extracted light curves for 8923 stars, including
a light curve for each of the RR Lyrae variables for which we have real light curve data.

\subsubsection{Results}
\label{sec:fake_results}

\begin{figure} 
\includegraphics[width=8.cm,height=8.cm]{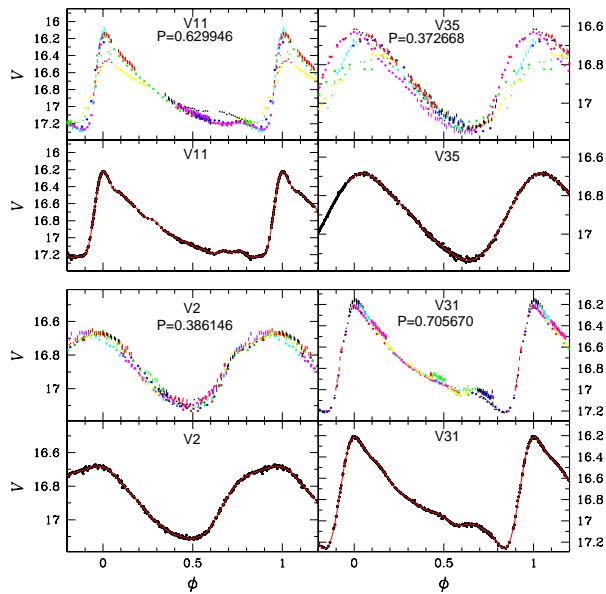}
\caption{ al light curves and their counterparts measured from the simulated images for
two clear Blazhko variables (V11 and V35) and two suspected cases (V2 and V31).
The stability of the light curves extracted from the simulated images demonstrates that the modulations in the observed light curves are real.
The red curves represent the fits to the real light curves that were used to generate the variable star fluxes in the simulated images.}
    \label{fakelc}
\end{figure}

In Figure~\ref{fakelc}, we plot the real phased light curves (upper panel in each case) and the light curves extracted from the simulated images
(black points in the lower panel in each case) for two RRab stars and for two RRc stars. V11 and V35 have very clear Blazhko effects in the real light curves, and
V2 and V31 have suspected low-amplitude Blazhko effects. The most striking feature of these plots is how closely the simulated light curves 
follow the input model light curves (red curves in the lower panel in each case) with
what seems to be only random scatter around the input model consistent with the data point uncertainties. To the eye, it is clear
that our reduction pipeline has not introduced any systematic errors that could be mistaken as a Blazhko effect, even in the 
less obvious cases of V2 and V31. This observation is true of all the light curves extracted from the simulated images.

However, to be able to analyse these results more robustly, we must find a way to quantify the amount of Blazhko effect that is present in a phased light
curve so that we may quantitatively compare our real light curves with the simulated light curves, and we do this in the next Section.

\subsection{A detection statistic for Blazhko modulations}

While most Blazhko modulated RR Lyrae variables can be distinguished by a close visual inspection of the phased light curve, the marginal cases may be subject
to the personal opinion of the observer. With the aim of quantifying the level of the Blazhko effect present in a light curve, and to
provide a numerical method for detecting its presence, we need to develop an appropriate detection statistic. We consider
the ``alarm'' $\cal A$ statistic designed by Tamuz, Mazeh \& North (2006) (see their Section 4 and Equation 3) for the detection of systematic
deviations of a light curve from a fitted model. In contrast to the $\chi^2$ goodness-of-fit statistic, the
Tamuz $\cal A$ statistic takes into account both the size of the residuals and the size of the runs of consectutive residuals
of the same sign. Such a statistic is therefore well-adapted to the detection of the 
Blazhko effect which can be considered as a systematic deviation from a stable perfectly-repeating light curve.

We found that clear Blazhko variables with large amplitude modulations actually produce a low value of the $\cal A$ statistic, and that
non-modulated stars tend to have relatively large $\cal A$ statistic values. This is contrary to what we expect and to the philosophy of the $\cal A$
statistic. We traced this problem down to the fact that the $\cal A$ statistic is normalised by the $\chi^2$ statistic. Blazhko variables
with large amplitude modulations have both long consecutive groups of large residuals of the same sign and a large $\chi^2$ value, which cancel out
in the calculation of the $\cal A$ statistic to produce an undesirably low value. We have modified the $\cal A$ statistic in order to 
avoid this design flaw by choosing to normalise the statistic by the number of data points rather than by the $\chi^2$.

We define a new statistic $\cal S_B$ for the detection of the Blazhko effect as follows:
\begin{equation}
{\cal S_B} = {1 \over N} \sum_{i=1}^{M}({r_{i,1}\over \sigma_{i,1}} + {r_{i,2}\over \sigma_{i,2}} + ... + {r_{i,k_i}\over \sigma_{i,k_i}})^2,
\label{eq:alarm}	
\end{equation}
\noindent
where $N$ is the total number of data points in the light curve, and $M$ is the number of groups of time-consecutive residuals of the same sign from 
a perfectly-repeating light curve model (e.g. a Fourier decomposition fit). The residuals $r_{i,1}$ to $r_{i,k_i}$ form the $i$th group of
$k_i$ time-consecutive residuals of the same sign with corresponding uncertainties $\sigma_{i,1}$ to $\sigma_{i,k_i}$. Our Blazhko detection
statistic $\cal S_B$ may therefore be interpreted as a measure of the systematic deviation per data point of the light curve from the perfectly-repeating model.

\begin{figure}
\begin{center}
\includegraphics[width=9.cm,height=9.cm]{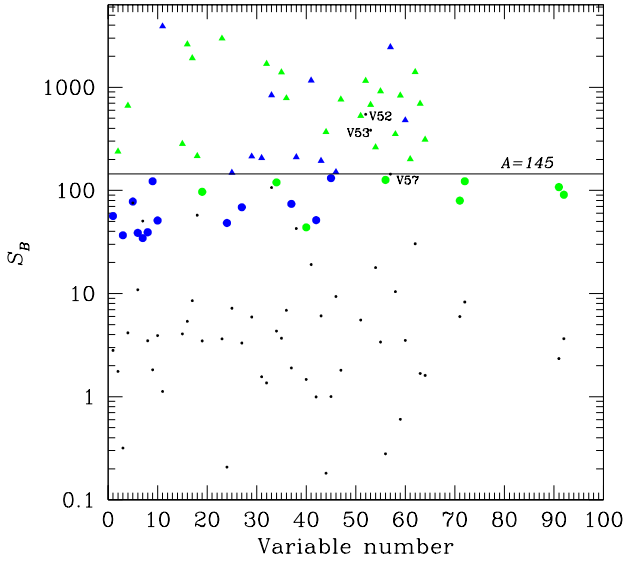}
\caption{The $\cal S_B$ statistic for all RR Lyrae stars. Blue symbols are for RRab stars, green symbols for RRc stars.
Triangles are used for modulated stars and filled circles for
non-modulated stars. The value of $\cal S_B$=145 is our adopted threshold line
between modulated and non-modulated stars.  Small black dots are used for values
from the simulated light curves, among which three peculiar cases are labeled: V52, V53 and V57. See text for a discussion.}
    \label{fig:alarm}
\end{center}
\end{figure}

Fig. \ref{fig:alarm} shows the values of the $\cal S_B$ statistic for the real light curves of the RR Lyrae stars plotted against the variable number.
Blue and green symbols are used for the RRab and RRc stars respectively. We also plot the values of the $\cal S_B$ statistic for the corresponding
light curves extracted from the simulated images in Section~\ref{sec:red_fake} as the small black dots. It is clear that the $\cal S_B$ statistic
values are much smaller in general for the simulated light curves when compared to the real light curves, which demonstrates that the
observed amplitude and/or phase modulations of the RR Lyrae stars are real, and not an artifact of the reduction process.

The two stars with by far the largest values of the $\cal S_B$ statistic for the simulated
light curves are V52 and V53, which lie in close spatial proximity to each other
(see the finding chart in Paper I). Since the {\it simulated}
light curves of these two variables show amplitude and phase variations, it is clear that they are suffering mutual flux contamination during the reduction process.
We have ignored these two stars in setting the threshold of the $\cal S_B$ statistic for Blazhko detection, and instead we set a threshold
of $\cal S_B$=145 just above the $\cal S_B$ statistic value for the simulated light curve of V57, which has the next largest value. This
threshold is plotted in Figure \ref{fig:alarm} as a horizontal line. We consider that we have detected the Blazhko effect in a real light curve
if its corresponding $\cal S_B$ statistic value is greater than this threshold, and we plot such variables with a triangle symbol. Stars for
which we do not detect the Blazhko effect are plotted as circles.

By adopting a threshold for the $\cal S_B$ statistic that lies above the $\cal S_B$ statistic values for all of our simulated light curves
(with the exception of the special cases of V52 and V53), we are ensuring that we would not erroneously ``detect'' the Blazhko effect in
any of the simulated light curves (a false positive detection). This is important since the values of the $\cal S_B$ statistic for the
simulated light curves include the contribution of the systematic errors introduced by our reduction software in our photometry, which will also be present
in our real light curves.

On the basis of the above discussions we list in Table \ref{tab:blazhko} the variables in NGC~5024 for which we have detected the
Blazhko effect. Stars with $\cal S_B <$145 are considered to be non-modulated stars within the limits of our data and are listed in Table \ref{tab:nonBl}.

Close to the threshold line in Fig. \ref {fig:alarm} there are some stars whose Blazhko nature will most likely be clarified
when more data of good quality and denser sampling become available.

\begin{table*}
\caption{Detected Blazhko variables in NGC 5024. 
The main frequency $f_0$ corresponds to the period 
reported in Table 2. Under $A_m$, we list the amplitude of the modulation displayed in the light curves of Figs. \ref{BLAZHKO_RR0_V} and \ref{BLAZHKO_RR1_V}. The $\cal S_B$  statistic  is calculated using
Eq. \ref{eq:alarm}; in this table we 
include only stars with $\cal S_B \geq$ 145. In the 
comments column, we indicate whether the star is located in the ``outer'' or ``inner'' region of 
the cluster and we provide some information on its blend status. All stars in this table are plotted with colour simbols in Figs. 1-4.}
         \begin{center}
\begin{tabular}{@{}lcccrl@{}}
\hline
Variable   &Bailey's &$f_0$ &$A_m$ &$\cal S_B$ & Comments \\
           & type    &(c/d) &(mag)&  &\\
\hline
 V2    &RRc &2.589694 &0.04&238.5& outer,nb\\
 V4    &RRc &2.593300 &0.030&662.5&outer,nb\\
 V11$^a$&RRab&1.587438&0.35&3902.3& outer,nb\\
 V15   &RRc &3.239475 &0.090&283.5& outer, BPix\\
 V16$^a$ &RRc &3.298588 &0.035 &2602.5& outer,nb\\
 V17   &RRc &2.623997 & 0.033 &1918.0& outer,nb\\
 V18   &RRc &2.975659 &0.045 &215.8& outer,nb\\
 V23   &RRc &2.731188 &0.075& 2972.7& outer,nb\\
 V25    &RRab &1.418132 &0.069 &148.7& outer,nb\\
 V29    &RRab &1.214699 &0.038 &213.8& outer,nb\\
 V31    &RRab &1.417093 &0.063 &206.6& border,nb\\
 V32    &RRc &2.563603 &0.057&1692.5&  outer,nb\\
 V33 &RRab&1.601052&0.02& 838.7& outer,blB\\
 V35    &RRc &2.683354 &0.200&1399.4& outer,nb\\
 V36    &RRc &2.678803 &0.035&784.4 & outer,nb\\
 V38    &RRab &1.416846 &0.029 &210.4& outer,nb\\
 V41   &RRab&1.627498 &  0.143 &1165.1& border,nb\\
 V43    &RRab &1.404467 &0.020 &193.4& inner,nb\\
 V44   &RRc &2.667321 & 0.062&369.3& inner,nb\\
 V46    &RRab &1.421163 &0.037 &151.0& inner,nb\\
 V47   &RRc &2.981817 & 0.060&761.2& outer,nb\\
 V51    &RRc &2.815196 & 0.080&529.7& inner, blB\\
 V52$^b$     &RRc &2.742574 &0.086 &1158.3& inner, close to V53\\
 V53$^b$     &RRc &2.570166 &0.083 &677.7& inner, close to V52\\
 V54    &RRc &3.173525&0.096 & 261.9 &  inner, blB\\
 V55 &RRc &2.256180&0.058&913.8&inner,nb\\
 V57$^a$ &RRab&1.759807 &0.393 &2455.7 & inner, blB\\
 V58 &RRc &2.817132 &0.093&351.7& inner, blB\\
 V59 &RRc &3.290101&0.077&832.1&inner,bl \\
 V60 &RRab&1.550974&0.105&479.2&inner,blB \\
 V61 &RRc &2.703214& &201.3 & inner,bl \\
 V62 &RRc &2.778650& 0.053&1404.9&inner,nb  \\
 V63 &RRc &3.220965 &0.078&694.6&  inner,nb\\
 V64&RRc &3.127698 &0.037&311.6& inner,nb\\
\hline
\end{tabular}
\end{center}
\label{tab:blazhko}
$^a$Announced as Blazhko variable by D\'ek\'any \& Kov\'acs (2009),$^b$ modulations may be an artifact due to proximity with another RR Lyrae\\
nb: not blended; bl: blended with fainter star; blB: blended with brighter star; Bpix: near to bad pixels
\end{table*}

\section{Discussion}
\label{sec:discussion}

Globular clusters rich in RR Lyrae stars often have large proportions of Blazhko variables, for
example M3 (Smith 1981), M5 (Jurcsik et al. 2011), M68 (Walker 1994), 
$\omega$ Cen (Jurcsik et al. 2001), NGC 362 (Sz\'ekely et al. 2006) and NGC 3201 
(Piersimoni et al. 2002).
It has been a generally accepted idea that the Blazhko effect occurred in 20-30\% of the RRab stars. 
However, the analysis of large data sets and detailed studies based on high-quality and long-term data
of individual stars, has revealed 
that the incidence of the Blazhko phenomenon may be as large as 50\% in RRab stars, and that amplitude and phase modulations 
may be present in hitherto unknown cases (e.g Jurcsik et al. 2009, 
Chadid et al. 2010; Poretti et al. 2010; Benk\H o et al. 2010).

\begin{figure} 
\includegraphics[width=8.0cm,height=12.0cm]{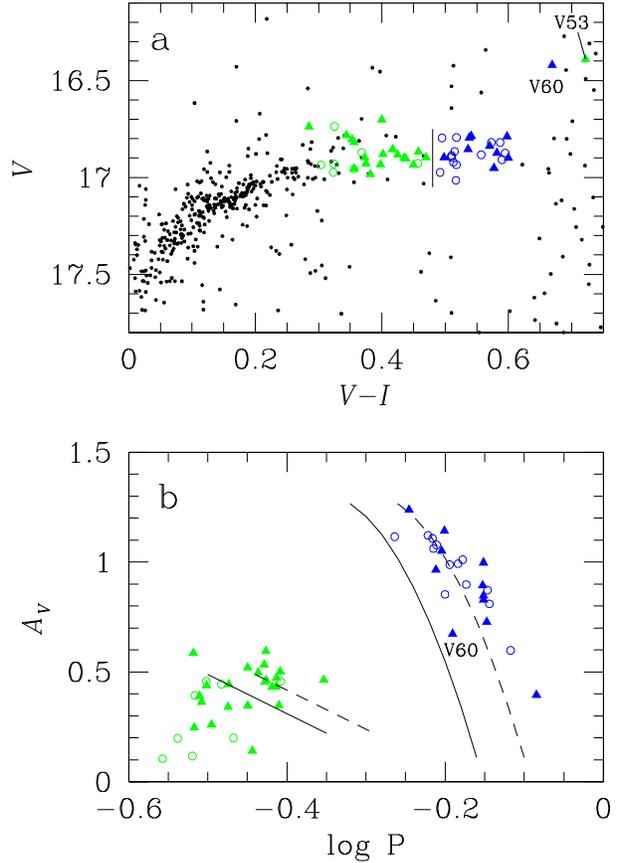}
\caption{Panel a) shows the distribution of the RR Lyrae stars in the 
Horizontal Branch. RRc stars are green symbols and RRab stars are blue. Filled triangles are used for stars for which we detect Blazhko modulations
and empty circles are used otherwise. 
Panel b) displays the Bailey diagram for NGC~5024. 
The continuous lines represent the average distribution of the RRab and RRc stars in the 
OoI cluster M3, and the segmented lines are the loci of 
evolved stars (Cacciari et al. 2005) often seen in OoII clusters like NGC~5024. 
See Section \ref{sec:discussion} for a discussion.}
    \label{TRENDS}
\end{figure}

\begin{figure}
\begin{center}
\includegraphics[scale=0.65]{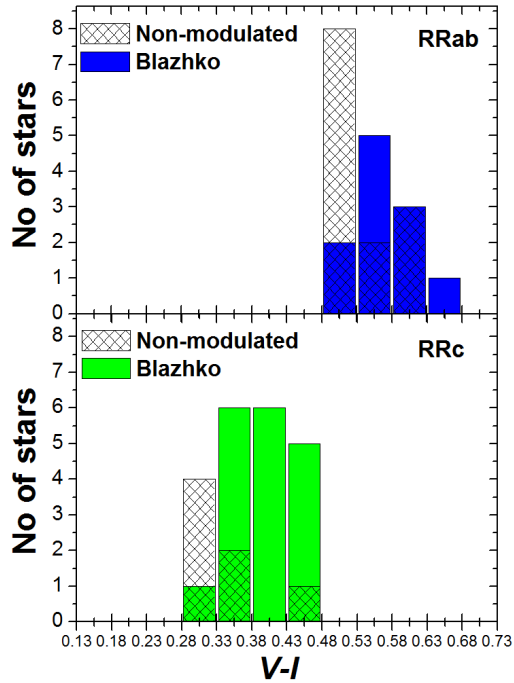}
\caption{Colour distribution of Blazhko variables (solid colour bars) and non-modulated stars (dashed bars). In average the Blazhko RRc stars are redder than 
their stable counterparts and tend occupy the either-or or hysteresis region, just to the blue of the fundamental pulsators region. Star V53 and V60 labeled in panel a of Fig. \ref{TRENDS} are not considered.}
    \label{histograms}
\end{center}
\end{figure}

\begin{table}
\caption{RR Lyrae stars classified as non-modulated on the basis of the $\cal S_B$ statistic ($\cal S_B <$ 145). Despite of having a low value of $\cal S_B$, in Figs. 1-4 we have 
plotted in colour those stars whose light curves display apparent amplitude modulations. }
         \begin{center}
\begin{tabular}{@{}lccl@{}}
\hline
Variable   &Bailey's &$\cal S_B$-\it{index}& Comments \\
           & type    &  &\\
\hline
V1    &RRab &56.43 &outer,nb  \\
V3    &RRab &36.7 & outer,nb \\
V5    &RRab & 78.0 & outer,nb \\
V6    &RRab &38.7 &outer,nb  \\
V7    &RRab &34.5& outer,nb \\
V8    &RRab &39.3 & outer,nb \\
V9    &RRab &122.7 & outer,nb \\
V10    &RRab & 51.1 &outer,nb  \\
V19$^a$    &RRc  &97.0 & outer,nb\\
V24    &RRab &48.4 &outer,nb  \\
V27    &RRab &68.7 &outer,nb  \\
V34$^a$   &RRc  &119.6 &outer,nb  \\
V37    &RRab &74.0 &border,nb  \\
V40$^a$   &RRc  &43.8 &outer,nb  \\
V42$^a$    &RRab &51.5 & border,nb\\
V45$^a$    &RRab &131.6& inner,nb\\
V56$^a$   &RRc  & 126.4 &inner,nb  \\
V71$^a$    &RRc  & 79.6& inner,blB\\
V72$^a$    &RRc  &122.7& inner,nb\\
V91$^a$    &RRc  &107.7& inner,nb\\
V92$^a$    &RRc  &90.8 & inner,nb\\
\hline
\end{tabular}
\end{center}
\label{tab:nonBl}
$^a$ suspected of amplitude modulations and plotted in colour in Figs. 1-4; nb: not blended; blB: blended with brighter star 
\end{table}

Most of the above discussions on the incidence of Blazhko variables have principally involved the
number of fundamental pulsators (or RRab stars) since the effect in first overtone pulsators 
(or RRc stars) has been detected only
in the last ten years and their number is too small for statistical work. Olech et al. (1999)
found the first evidence of modulating non-radial pulsations in three RRc stars in M55. Alcock et al. (2000) reported frequency sidepeaks of main frequency in 2-4\% of the RRc variables in the LMC from the MACHO data.
Jurcsik et al. (2001) detected amplitude
modulations in 8 of 48 RRc stars in $\omega$~Cen. According to Piersimoni et al. (2002), the RRc star V48
in NGC 3201 shows amplitude modulations from a single comparison of their light curve amplitude 
with the amplitude reported by Sawyer-Hogg (1973). However, no evidence is given of an
amplitude and/or phase modulated light curve. From the OGLE-I sample, Moskalik \& Poretti (2003) 
identified modulation frequencies in 3 RRc stars in the Galactic bulge (see their Table 8).

In their recent work on M5, Jurcsik et al. (2011) compare period and colour distributions of 
Blazhko variables and stable RR Lyrae stars. No RRc stars with the 
Blazhko effect are known in M5. The Blazhko variables discovered in this 
paper bring the number to 23 confirmed cases among the RRc stars which, 
to the best of our knowledge, makes NGC~5024 the globular cluster with the largest number 
of RRc Blazhko variables. Hence, it is of interest to see if the new large population of Blazhko RRc
variables in NGC~5024 is consistent with the trends found by Jurcsik et al. (2011) in M5.

\subsection{Trends in brightness, colour and amplitudes of Blazhko variables}
\label{sec:trends}

Panel ($a$) of Fig. \ref{TRENDS} shows the horizontal branch and the distribution of the Blazhko 
and stable 
RR Lyrae stars. There is a clear separation of the RRab and RRc stars indicated by the vertical solid line at $V-I = 0.48$. This has already been noted in Paper I 
and has been interpreted as the border between the either-or-region 
(or region where due to hysteresis the first overtone and fundamental pulsators might coexist) (Bono et al. 1997; Szabo et al. 2004)
and the fundamental (F) region.
The clean separation of fundamental and first overtone pulsators
is probably produced when stellar evolution occurs 
towards the red (Caputo et al. 1978). The two labelled stars, V53 and V60, already
noticed and discussed in Paper I, are ignored since they
have a peculiar position in the CMD likely due to contamination from a blended star. 
The RRc stars with Blazhko modulations (filled triangles) in average are redder
than the non-modulated stars.

In panel ($b$) of Fig. \ref{TRENDS}, we present a Bailey diagram where one can see that for the RRab stars, the 
Blazhko and non-Blazhko variables are well mixed, and thus we do not see
the trend noticed by Jurcsik et al. (2011) for the case of M5
that Blazhko stars have larger amplitudes and shorter periods. 
For the RRc stars there is an indication that the Blazhko variables have
larger amplitudes and periods but the distribution is very scattered,
which is contrary to
the trend found in non-modulated stars in M3 or other OoII type clusters (Cacciari et al. 2005). 
The scatter among the Blazhko RRc stars in NGC 5024 (green triangles in Fig. \ref{TRENDS}) is most likely connected with amplitude modulations. The dispersion among the non-modulated RRc (green open circles in Fig. \ref{TRENDS}), 
of shorter period and smaller amplitude, reminds the equally scattered distribution of the short-period small-amplitude RRc variables V105, V178, and V203 in M3 (see Fig. 2 of Cacciari et al. 2005).

The $V-I$ colour distribution of the RR Lyrae stars in NGC~5024 can also appreciated in the histogram of Fig. \ref{histograms}. 
The distribution of the Blazhko RRc 
variables is redder than that of the stable RRc. This, along with the indication that the evolution is towards the red, may suggest that the Blazhko effect
manifest systematically in stars undergoing a change in pulsation mode.

\subsection{Times of maximum brightness}
\label{sec:TIMAX}

The times of maximum brightness in a variable star are of fundamental importance since a
long record of them help to trace evolutionary changes in the star. In the case of 
Blazhko variables the variations of the maximum light bear information on the periodicity
of the modulation cycle and help to trace the long term variations of the cycle. Dense 
records of times of maximum is challenging as it requires extensive timely observations.
With the conviction that in the future the Blazhko variables in NGC~5024 will be
further observed and studied, we measured as many times of maximum light
as it has been possible in the light curves of Figs. \ref{BLAZHKO_RR0_V} and \ref{BLAZHKO_RR1_V} and have kept a record of them in Table \ref{tab:timax-e}, of which
only a small portion is given in the printed version of this paper but the full table is available in electronic form. We have been able to measure between 5 and 11 times of maximum light per star for the majority of the stars. Since the time distribution of our data set
is rather limited, we have refrain from attempting a determination of the
periodicity in these times of maximum light,
but we trust that they will be useful in the future.

\begin{figure*}
\begin{center}
\includegraphics[width=17.5cm,height=10.cm]{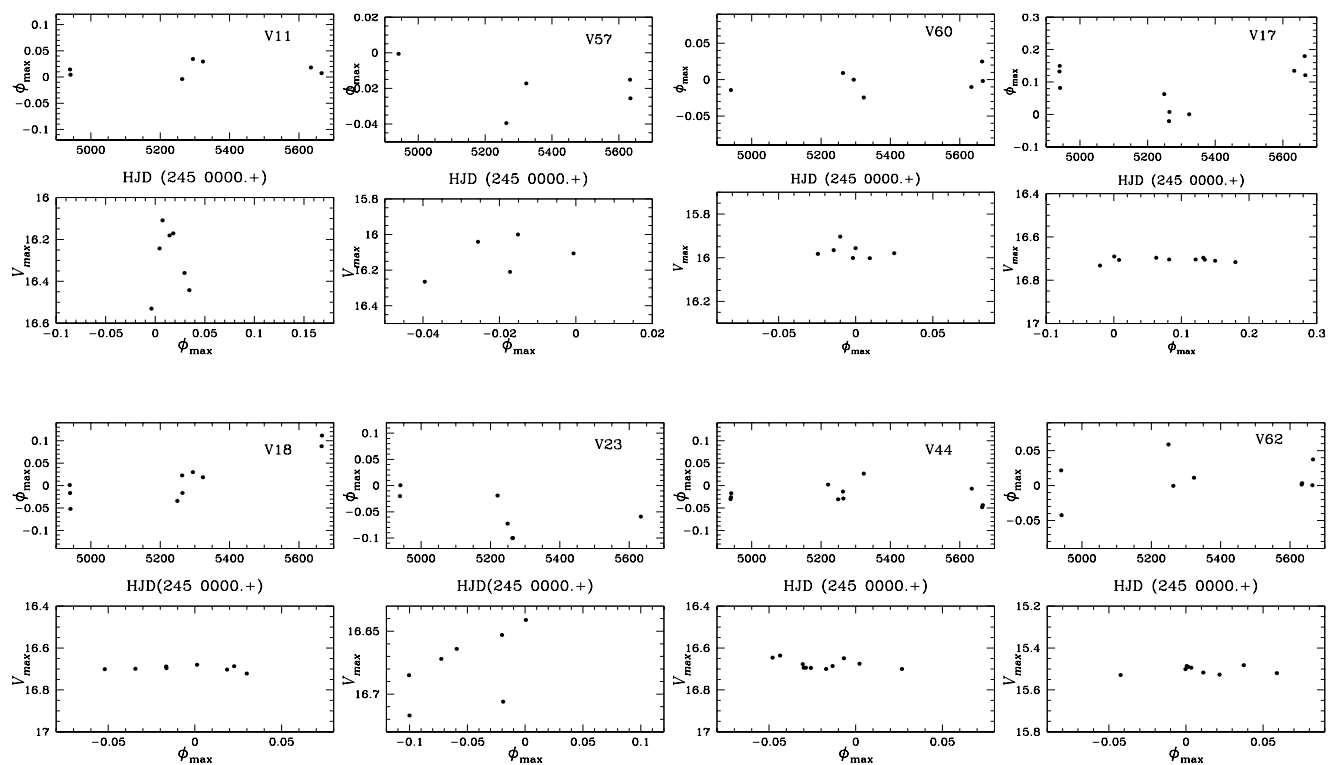}
\caption{Phase and maximum brightness modulations in for three selected RRab stars, V11, V57 and V60 and five
RRc stars V17, V18, V23, V44, V62.}
    \label{fig:MAXIMA2cajas}
\end{center}
\end{figure*}

In the upper panels of Fig. \ref{fig:MAXIMA2cajas}, the phase variations are shown for a group
of selected stars with well determined times of maximum light. If the phases on the vertical axis are multiplied by the pulsation period, these plots correspond to the $O-C$ diagrams showing the period variations produced by the Blazhko modulations. The bottom panels display the "loop" diagrams which show the correlation between the amplitude and phase modulations. In cases of 
continuous dense observations of Blazhko variables (e.g. Guggenberger et al. 2011; Chadid et al. 2010a,b) these plots 
show loops that indicate whether or not the amplitude and phase modulations are taking place in phase with each other. In our case, despite the sparseness of the Blazhko 
phase coverage, the plots show general trends between the maximum brightness and phase  modulations; we note that for some stars the diagrams 
display clear traces of the loops (e.g. V11, V57, V23) whereas some others
show mild linear trends (V44) or phase variations at constant maximum brightness (V17, V18, V62).

\begin{table*}
\caption{Phase and magnitude at the time of maximum brightness in the $V$-band for a group 
of stars with Blazhko modulations. The phase in column 2 was calculated using the ephemerides given in Table \ref{variables}. This is an extract from the full table, which is available with the electronic version of the article (see Supporting Information).}
\centering
 \begin{tabular}{cccc}
\hline
 Star & $\phi_{max}$ & $V_{max}$ &  HJD$_{max}$      \\
    &       &      & (245 0000.+)\\
\hline
V2  &+0.000 &  16.656    & 4940.34212\\
V2  &-0.039 &  16.690    & 5249.24358\\
... & ...   &   ...      &    ...    \\
V4  &+0.000 &  16.681    & 4939.27817\\
V4  &-0.004 &  16.688    & 4941.20439\\
... & ...   &   ...      &    ...    \\
V11 & +0.014&  16.181    & 4940.20614\\
V11 & +0.004&  16.243    & 4941.45969\\
... & ...   &   ...      &    ...    \\
\hline
\end{tabular}
\label{tab:timax-e}
\end{table*}

\section{Conclusions}

Precise, dense, and long time-span time-series photometry is required to detect amplitude and phase modulations in RR Lyrae variables. If this is achieved, then one may find that the Blazhko effect 
is more common than generally realized, particularly among the first overtone pulsators or RRc stars.
The significant increase in the detections of Blazhko variables in NGC~5024 is mainly due to 
the high quality of the CCD photometry achieved using the DIA technique which enables one
to detect small amplitude and phase modulations. The accuracy of our $V$ photometry at the brightness 
of the RR Lyrae stars is about 0.008 mag per data point, which is, in most of the confirmed Blazhko variables, significantly smaller than 
the amplitude of the modulations in the light curves.

Our data reveal that amplitude and phase modulations are present in the majority of the light curves of the RRc
stars (23 of 31) and in about half of the RRab stars (11 of 24) in NGC 5024, i.e. 66\% and 37\% respectively. We also find a lower limit of 52\% on the overall incidence rate of the Blazhko effect among
the RR Lyrae population of the cluster, a fraction which may increase as more extensive 
time-series CCD photometry becomes available.

It is likely times of maximum O-C residuals observed in Blazhko variables are 
caused by the phase modulations. Given the short time base of our observations we cannot
say whether or not these O-C differences are at all connected with secular variations in the pulsation period.

The RRab and RRc stars are cleanly separated in the CMD which is the expected situation 
if the evolution of the RR Lyrae stars is progressing towards the red.
In the prevailing hysteresis paradigm, the current pulsation mode of an RR Lyrae 
star is defined by its
previous evolutionary history (e.g., van Albada \& Baker 1973; Caputo
et al. 1978). As schematically shown in Fig. 3 of Caputo et al. (1978),
one only expects to find a clean separation between RRab and RRc stars if horizontal
branch evolution is predominantly towards the red in the instability strip.
At the same time, the color distribution of the Blazhko RRc variables
is redder than that of the stable counterparts, and populate the hysteresis region 
where first overtone and fundamental pulsators may coexist. This may suggest
that the Blazhko effect manifests in stars undergoing a pulsation mode change.

Our limited time coverage of the Blazhko cycle prevented us from estimating 
the modulation periods. Despite of this the light curve modulations could be
 detected with confidence. A continued monitoring of NGC~5024 should prove
useful in characterizing the modulation properties of the confirmed and suspected Blazhko stars, in particular the determination of the modulation 
frequencies. The large sample of Blazhko variables in NGC~5024 will eventually
be used to confirm the existance of the correlation between the main pulsation frequency $f_0$ and the modulation frequency $f_m$, sketched for the Blazhko RRab stars in M5 (Jurcsik et al. 2011) and whether the zero point is metallicity dependent. Comprehensive studies of the Blazhko-modulated stars 
in globular clusters might enable us to study metallicity and stellar evolution effects
in RR Lyrae stars and infer incidence rates of the Blazhko modulation in
homogeneous populations.

\section*{Acknowledgments}

We are greatful to the referee for multiple constructive
comments and suggestions. AAF is thankful to the Indian Institute of
Astrophysics for hospitality during his sabbatical leave in 2010 and acknowledges support of 
CONACyT (Mexico), DST (India) and 
from DGAPA-UNAM through grant IN114309. This work has made a large use 
of the SIMBAD and ADS services.

\end{document}